\documentclass[9pt,twocolumn,twoside]{opticajnl}
\journal{opticajournal} % use for journal or Optica Open submissions

\usepackage{graphicx}
\usepackage{caption}
\setboolean{shortarticle}{true}
\usepackage{lineno}
\title{Unveiling the Relationship Between Amplitude-only Transmission Matrix and Third-Order Correlation of Light Fields in Wavefront Shaping}

\author[1]{Feixiang Ren}
\author[1,2,*]{Haoyi Zuo}

\affil[1]{College of Physics, Sichuan University, Chengdu, Sichuan 610064, China}
\affil[2]{School of Science, Chongqing University of Technology, Chongqing 400054, China}
\affil[*]{zuohaoyi@scu.edu.cn}

\begin{abstract}
In the regime of wavefront shaping (WFS) techniques, both the transmission matrix (TM) methods and a recently proposed third-order correlation of light fields (TCLF) method are effective in overcoming light scattering. This letter details the relationship between the amplitude-only TM method and the amplitude-only TCLF. The random fluctuations of different pixels on a digital micromirror device (DMD) in the amplitude-only TCLF method can be regarded as orthogonal bases, sharing the same role as the Hadamard bases in the amplitude-only TM method. This insight explains why the computational complexity of the TCLF method is significantly lower than that of the TM methods and also indicates that the amplitude-only TM is essentially a special case of the TCLF method.
\end{abstract}

\setboolean{displaycopyright}{false} % Do not include copyright or licensing information in submission.

\begin{document}

\maketitle
Scattering has consistently posed a significant challenge for optical imaging. To address it, many wavefront shaping (WFS) techniques have been proposed, such as the optical phase conjugation (OPC) \cite{ma2014time,liu2017focusing,yu2019implementation} and transmission matrix (TM) methods \cite{tao2015high,xu2017focusing,popoff2009measuring,akbulut2011focusing}. Among these, TM methods are widely used due to their clear physical principles and straightforward experimental setups. The transmission matrix $K$ contains the mapping relationship between the input and output light fields. The basic idea is to load a set of orthogonal input bases onto the spatial light modulator (SLM) and measure the corresponding output light intensities on the CMOS plane. $K$ will be obtained by matrix multiplication on the computer. The SLM can be a digital micro-mirror device (DMD) or a liquid-crystal SLM (LC-SLM),  which performs amplitude or phase modulation of the input plane wave, respectively. Thus the TM methods can be divided into two classes: the amplitude-only TM and the phase-only TM. The phase-only TM method \cite{popoff2009measuring} was first proposed using a four-phase method to obtain output light fields. Another research \cite{tao2015high} demonstrated an amplitude-only TM method by loading Hadamard bases onto the DMD along with a reference light. Typically only a small part of the pixels on the SLM, such as $64 \times 64$ pixels, is used in experiments. Using too many pixels would result in computer memory overflow \cite{yu2017ultrahigh}.

\begin{figure}[htbp]
\centering
\includegraphics[width=\linewidth]{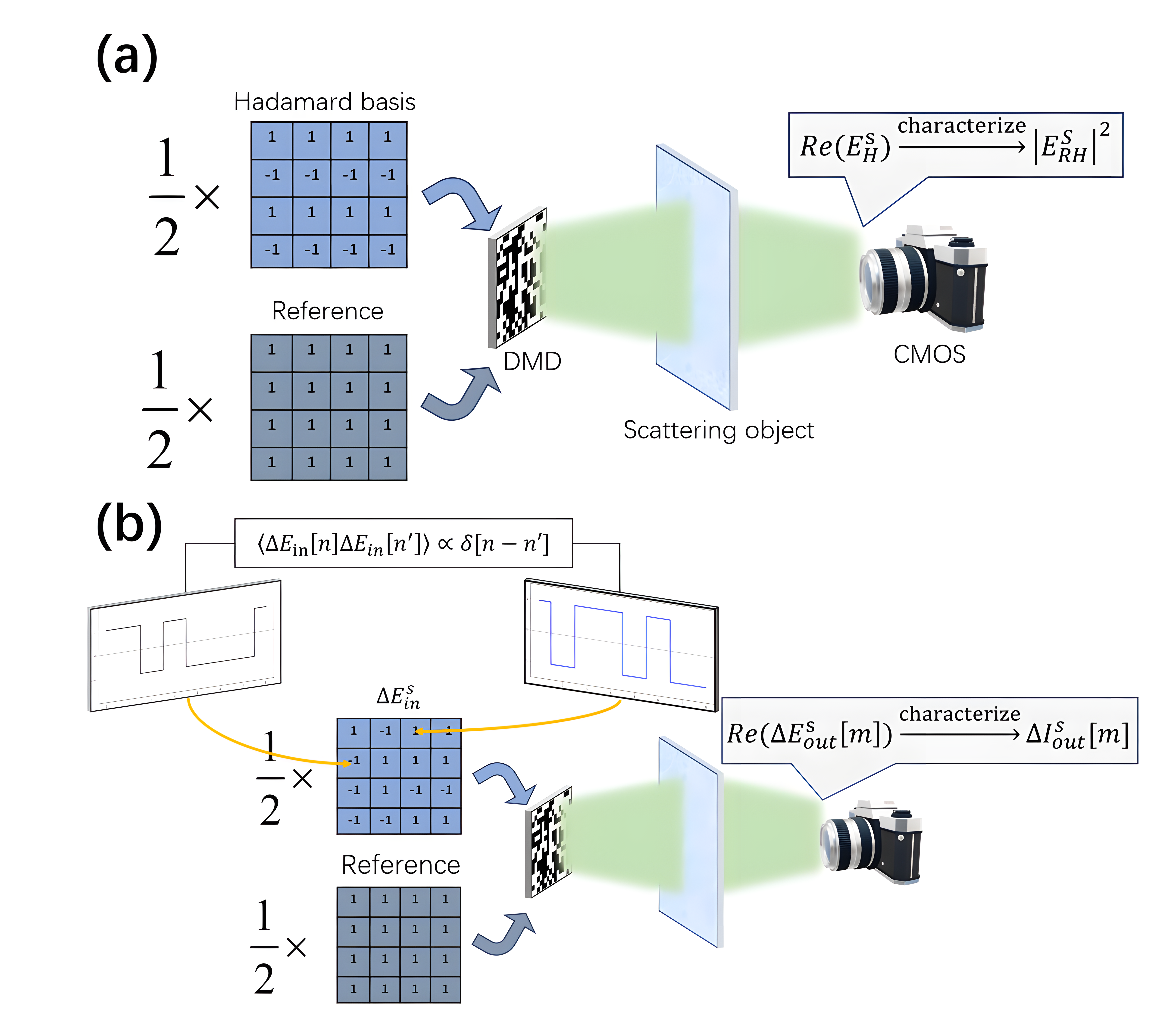}
\caption{(a)The schematic diagram of the amplitude-only TM method. The Hadamard bases are loaded onto the DMD along with the reference light $A_\text{ref}$. (b)The schematic diagram of the amplitude-only TCLF method re-interpreted with matrix theory. Both methods characterize the real part of light fields with intensity information that can be recorded by a CMOS camera.}
\label{picture1}
\end{figure}

Until recently, a low-cost WFS technique called the third-order correlation of light fields (TCLF) \cite{zhao2023low} was proposed by exploiting the statistical properties of fluctuating light fields. The concept originates from ghost imaging (GI) \cite{bromberg2009ghost,bennink2004quantum,ferri2005high,cheng2004incoherent}. Typically, GI correlates the input and output light intensities' fluctuations to obtain objects' amplitude information. In contrast, TCLF retains the phase information of the scattering object by correlating the fluctuations of the input light fields (instead of the light intensities) and output light intensities. In the TCLF method, a set of patterns that follow a circular complex Gaussian random process with a non-zero mean are loaded onto the SLM, and the corresponding output intensities are measured on the CMOS plane, as shown in Fig. \ref{picture1}(b). The TCLF pattern can then be obtained by numerical calculation and reloaded onto the SLM, forming a focus on the CMOS plane. Notably, full pixels on the LC-SLM ($1,920 \times 1,080$) can be used in the TCLF method, demonstrating its ability to mitigate the heavy burden on PC memory that all TM methods cannot avoid. In addition, the TCLF can also be achieved by both DMD and LC-SLM, categorizing it into an amplitude-only TCLF method and a phase-only one.

In this letter, we explore the relationship between the amplitude-only TM and the amplitude-only TCLF method, offering a theoretical explanation of the difference in their requirements for PC memory. In the last part, we will demonstrate that the amplitude-only TM method is essentially a special case of the amplitude-only TCLF method. We begin by detailing the theoretical relationship between the $K$ matrix and the TCLF pattern from the perspective of matrix theory. ${{E}_{\text{out}}}=K{{E}_{\text{in}}}$ is the key to all TM methods, where input and output light fields are discretized into vectors. By substituting this equation into the TCLF expression, we get
\begin{equation}
    \begin{aligned}
        \left\langle \Delta {{E}_{\text{in}}}[n]\Delta {{I}_{\text{out}}}[m] \right\rangle &=\left\langle {{E}_{\text{in}}}[n]{{I}_\text{out}}[m] \right\rangle -\left\langle {{E}_{\text{in}}}[n] \right\rangle \left\langle {{I}_\text{out}}[m] \right\rangle \\
        &=\sum\limits_{i}{\sum\limits_{j}{{{k}_{mi}}k_{mj}^{*}}} \left\langle {{E}_\text{in}}[n]{{{E}}_\text{in}}[i]E_\text{in}^{*}[j] \right\rangle \\
        &-\sum\limits_{i}{\sum\limits_{j}{{{k}_{mi}}k_{mj}^{*}}} \left\langle {{{E}}_\text{in}}[i]E_\text{in}^{*}[j] \right\rangle\left\langle {{E}_\text{in}}[n] \right\rangle, \\
    \end{aligned}
    \label{equation1}
\end{equation}
where $\left\langle...\right\rangle=\frac{1}{S}\sum\limits_{s}{...}$ denotes an ensemble average operation and $S$ is the sampling number, $\Delta {{E}_{\text{in}}}[n]={{E}_{\text{in}}}[n]- \left\langle {{E}_{\text{in}}}[n]\right\rangle$ denotes the fluctuations of the input light fields at the $n^\text{th}$ pixel, while $\Delta {{I}_{\text{out}}}[m]={{I}_{\text{out}}}[m]- \left\langle {{I}_{\text{out}}}[m]\right\rangle$ denotes the corresponding fluctuations of the output intensities at the $m^\text{th}$ pixel, where the focus would be located when reloading the TCLF pattern on the SLM.
According to the statistical property of a circular complex Gaussian random process with a non-zero mean, the third-order correlation of input light fields (not the TCLF expression) can be expressed in several terms of second-order correlations \cite{goodman2015statistical}, as shown below
\begin{equation}
    \begin{aligned}
        &\left\langle E_\text{in}[n] E_\text{in}[i] E_\text{in}^*[j]\right\rangle
        \\ &\qquad= \left\langle E_\text{in}[n] E_\text{in}[i]\right\rangle\left\langle E_\text{in}^*[j]\right\rangle+\left\langle E_\text{in}[n] E_\text{in}^*[j]\right\rangle\left\langle E_\text{in}[i]\right\rangle \\
        &\qquad+\left\langle E_\text{in}[i] E_\text{in}^*[j]\right\rangle\left\langle E_\text{in}[n]\right\rangle-2\left\langle E_\text{in}[n]\right\rangle\left\langle E_\text{in}[i]\right\rangle\left\langle E_\text{in}^*[j]\right\rangle.
        \end{aligned}
    \label{equation2}
\end{equation}
The second-order correlation of the fluctuations of the input light fields can be expressed as
\begin{equation}
    \begin{aligned}
        \left\langle \Delta E_\text{in}[n] \Delta E_\text{in}[n']\right\rangle=\left\langle E_\text{in}[n] E_\text{in}[n']\right\rangle-\left\langle E_\text{in}[n]\right\rangle \left\langle E_\text{in}[n']\right\rangle.
    \end{aligned}
    \label{equation2.1}
\end{equation}
Moving the term $\left\langle E_\text{in}[i] E_\text{in}^*[j]\right\rangle\left\langle E_\text{in}[n]\right\rangle$ in Eq. \ref{equation2} to the left side and substituting Eq. \ref{equation2.1} back to Eq. \ref{equation2} and, we can get
\begin{equation}
    \begin{aligned}
        &\left\langle E_\text{in}[n] E_\text{in}[i] E_\text{in}^*[j]\right\rangle - \left\langle E_\text{in}[i] E_\text{in}^*[j]\right\rangle\left\langle E_\text{in}[n]\right\rangle \\
        &\quad=\left\langle \Delta E_\text{in}[n] \Delta E_\text{in}[i]\right\rangle\left\langle E_\text{in}^*[j]\right\rangle+\left\langle \Delta E_\text{in}[n] \Delta E_\text{in}^*[j]\right\rangle\left\langle E_\text{in}[i]\right\rangle.
    \end{aligned}
    \label{equation3}
\end{equation}
Since the fluctuations on different pixels of the SLM represent the incoherence part of the light source, the second-order correlation of the fluctuation of the input light fields should be expressed as a delta function
\begin{equation}
    \begin{aligned}
        \left\langle \Delta {{E}_\text{in}}[n]\Delta {{E}_\text{in}}[n'] \right\rangle &=\frac{1}{S}\sum\limits_{s}{\Delta E_\text{in}^{s}[n]\Delta E_\text{in}^{s}[n']}\\
        &=\left\langle \Delta {{E}_\text{in}}[n]\Delta {{E}_\text{in}}[n] \right\rangle \cdot \delta [n-n'],
    \end{aligned}
    \label{equation4}
\end{equation}
where $E_\text{in}^{s}$ denotes the input light field imposed on the DMD at the $s^\text{th}$ sampling. By substituting Eq. \ref{equation3} and \ref{equation4} back to Eq. \ref{equation1}, the TCLF expression can be written as
\begin{equation}
    \begin{aligned}
        \left\langle\Delta E_{\text {in }}[n] \Delta I_{\text {out }}[m]\right\rangle= & \left\langle\Delta E_{\text {in }}[n] \Delta E_{\text {in }}[n]\right\rangle\left\langle E_{\text {out }}^*[m]\right\rangle k_{m n}+ \\
        & \left\langle\Delta E_{\text {in }}[n] \Delta E_{\text {in }}^*[n]\right\rangle\left\langle E_{\text {out }}[m]\right\rangle k_{m n}^*,\\
        \end{aligned}
        \label{equation5}
\end{equation}
for $n=1,2,..., N$, where N denotes the number of pixels on the DMD. The TCLF expression between the light field fluctuation at each pixel of the DMD and the intensity fluctuation at the $m^\text{th}$ CMOS pixel is computed simultaneously, yielding a TCLF vector which is then binarized to generate a TCLF pattern. $k_{m n}^*$ in the second term of Eq. \ref{equation5} offsets the phase distortion caused by the scattering object and focuses the light field at the $m^\text{th}$ pixel of the CMOS camera. 
%The above conclusion in Eq. \ref{equation5} applies to both the amplitude-only and the phase-only TCLF method. In the phase-only TCLF case, $k_{m n}^*$ in the second term of Eq. \ref{equation5} functions similarly to $K^{\dagger}E^\text{target}$ in the phase-only TM method \cite{popoff2009measuring}, where $\dagger$ represents conjugate transpose and $E^\text{target}$ denotes a vector with all elements 0 except for the $m^\text{th}$ position, which is set to 1. $k_{m n}^*$ for $n=1,2,...,N$ offsets the phase distortion caused by the scattering object and focuses the light field at the $m^\text{th}$ pixel of the CMOS camera as $K^{\dagger}E^\text{target}$ dose in the phase-only TM method. This conclusion is not surprising, as experiments have already demonstrated that both methods yield similar results: when reloading the phase-only TCLF pattern and the phase information of $K^{\dagger}E^\text{target}$ on the input plane respectively, a focus can form on the CMOS plane in both cases. 

%However, we can draw a more in-depth conclusion when comparing the amplitude-only TM and TCLF methods. 
With Eq. \ref{equation5}, we are prepared to proceed with the comparison between the amplitude-only TM and TCLF methods.
To help better illustrate the idea, a quick review of the theory behind the amplitude-only TM method \cite{tao2015high} is necessary. The schematic diagram of the amplitude-only TM is shown in Fig. \ref{picture1}(a). The principle of this method is
\begin{equation}
    \begin{aligned}
        &\operatorname{Re}\{\frac{E_\text{ref}^{*}[m]}{\left| {{E}_\text{ref}[m]} \right|}\left[ \begin{matrix}
        E_\text{H}^{1}[m] & ... & E_\text{H}^{S}[m]  \\
         \end{matrix} \right]\} \\& \qquad \qquad \qquad \qquad =\operatorname{Re}(\frac{E_\text{ref}^{*}[m]}{\left| {{E}_\text{ref}[m]} \right|} K_m)\left[ \begin{matrix}
         {{H}_{1}} & ... & {{H}_{S}}  \\
         \end{matrix} \right],
         \label{equation6}
    \end{aligned}
\end{equation}
where $K_m=\left[ \begin{matrix} k_{m 1} & ... & k_{m N} \end{matrix} \right]$ represents the $m^\text{th}$ row of the matrix $K$. ${H}_{s}$ denotes the $s^\text{th}$ column of a Hadamard matrix and $E_{H}^{s}[m]=K_m {H}_{s}$ is the corresponding output light field at the $m^\text{th}$ pixel of the CMOS camera when imposing ${H}_{s}$ on the input plane. $E_\text{ref}[m]=K_m A_\text{ref}$ is the output light field at pixel $m$ that we refer to as a (output) reference light if all the channels on the DMD are open, where $A_\text{ref}$ is an all-ones vector. Multiplying ${E}_\text{ref}^{*}[m]/{\left|{E}_\text{ref}[m]\right|}$ to all elements in $K_m$ represents a rotation operation on the complex plane to make sure that the reference light $E_\text{ref}[m]$ lies on the real axis (the original paper \cite{tao2015high} does not mention the rotation operation since it assumes that $E_\text{ref}[m]$ is already on the real axis). Extracting the real part of the rotated $K_m$, namely $\operatorname{Re}(\frac{E_\text{ref}^{*}[m]}{\left| {{E}_\text{ref}[m]} \right|} K_m)$ on the right side of equation \ref{equation6}, is to select the constructive elements in $K_m$ with phases in the range of $(\varphi_\text{ref}-\pi/2,\varphi_\text{ref}+\pi/2)$, where $\varphi_\text{ref}$ denotes the phase of $E_\text{ref}[m]$. However, the real part information of a light field, namely $\operatorname{Re}(\frac{E_\text{ref}^{*}[m]}{\left| {{E}_\text{ref}[m]} \right|}E_\text{H}^{s}[m])$, cannot be recorded by CMOS cameras. Thus, the (input) reference light $A_\text{ref}$ is loaded onto the DMD along with the Hadamard basis $H_{s}$ to characterize the real part of the light field with light intensity $I_\text{RH}^{s}[m]$, as shown below
\begin{equation}
	\operatorname{Re}(\frac{E_\text{ref}^{*}[m]}{\left| {{E}_\text{ref}[m]} \right|}E_\text{H}^{s}[m])\simeq \beta \left( \frac{ I_\text{RH}^{s}[m]}{\left| {{E}_\text{ref}[m]} \right|^2}-1 \right), 
    \label{equation7}
\end{equation}
where $\beta$ is a coefficient and $I_\text{RH}^{s}[m]=\left|E_\text{H}^{s}[m]+E_\text{ref}[m]\right|^2$ denotes the output light intensity recorded by the CMOS when imposing both a Hadamard basis and the reference light $A_\text{ref}$ on the DMD. After recording $I_\text{RH}^{s}[m]$ on the CMOS camera, $\operatorname{Re}(\frac{E_\text{ref}^{*}[m]}{\left| {{E}_\text{ref}[m]} \right|} K_m)$ can be computed through
\begin{equation}
    \begin{aligned}
        \operatorname{Re}(\frac{E_\text{ref}^{*}[m]}{\left| {{E}_\text{ref}[m]} \right|} K_m) \simeq \frac{1}{\gamma} \left[ \begin{matrix}
        I_\text{RH}^{1}[m] & ... & I_\text{RH}^{S}[m]  \\
         \end{matrix} \right] \left[ \begin{matrix}
         {{H}_{1}} & ... & {{H}_{S}}  \\
         \end{matrix} \right]^{T},
    \label{equation7.1}
    \end{aligned}
\end{equation}
where $\gamma$ is a constant. Binarizing it into a binary TM pattern and reloading the pattern onto the DMD, a focus can form on the CMOS plane.
\begin{table}[htbp]
\centering
\caption{\bf A Comparison of Similar Components in Two Amplitude-only Methods }
\begin{tabular}{ccc}
\hline
Components & TM & TCLF \\
\hline
Reference(input) & $A_\text{ref}$ & $\left\langle E_\text{in}\right\rangle$ \\
Reference(output) & $E_\text{ref}[m]$ & $\left\langle E_\text{out}[m]\right\rangle$ \\
Orthogonal bases & $H_{s}$ & $\Delta E_\text{in}^{s}$ \\
Input light fields & $\frac{1}{2}(A_\text{ref}+H_{s})$ & $E_\text{in}^{s}=\Delta E_\text{in}^{s}+\left\langle E_\text{in}\right\rangle$ \\
Output light fields & $I_\text{RH}^{s}[m]$ & $I_\text{out}^{s}[m]$ \\
\hline
\end{tabular}
  \label{tabel1}
\end{table}

The amplitude-only TCLF can also be interpreted using this paradigm, as shown in Fig. \ref{picture1}(b). In table \ref{tabel1}, we list the components that serve similar roles in these two methods. In the amplitude-only TM, Hadamard bases are chosen as input orthogonal light fields to measure all the elements in $K_m$. According to Eq. \ref{equation4}, $\Delta E_\text{in}^s[n]$ also satisfies the orthogonality. Thus we can select $\Delta E_\text{in}^s[n]$ as orthogonal bases, leading to the following equation
\begin{equation}
    \begin{aligned}
        &\left[ \begin{matrix}
            \Delta E_\text{out}^{1}[m]  & ... & \Delta E_\text{out}^{S}[m]  \\
         \end{matrix} \right] = \\ 
       &\qquad \quad \left[ \begin{matrix}
        {{k}_{m1}} & {{k}_{m2}} &... & {{k}_{mN}}  \\
     \end{matrix} \right]{{\left[ \begin{matrix}
         \Delta E_\text{in}^{1}[1] &  ... & \Delta E_\text{in}^{S}[1]  \\
         \Delta E_\text{in}^{1}[2] &  ... & \Delta E_\text{in}^{S}[2]  \\
         ... & ... & ...  \\
         \Delta E_\text{in}^{1}[N] & ... & \Delta E_\text{in}^{S}[N]  \\
      \end{matrix} \right]}}.\\ 
      \end{aligned} 
      \label{equation8}      
\end{equation}
By rotating the complex plane and extracting the real part, Eq. \ref{equation8} can be expressed as
\begin{equation}
    \begin{aligned}
        &\operatorname{Re}\{\left\langle E_\text{out}^{*}[m] \right\rangle\cdot\left[ \begin{matrix}
            \Delta E_\text{out}^{1}[m]  & ... & \Delta E_\text{out}^{S}[m]  \\
         \end{matrix} \right]\} = \\ 
       &\qquad\quad \operatorname{Re}\{\left\langle E_\text{out}^{*}[m] \right\rangle\cdot\left[ \begin{matrix}
        {{k}_{m1}} & {{k}_{m2}} &... & {{k}_{mN}}  \\
     \end{matrix} \right]\}\cdot \mathbf{\Delta E_\textbf{in}}, \\ 
    \end{aligned} 
    \label{equation9}      
\end{equation}
where
\begin{equation}
    \left\langle E_\text{out}[m] \right\rangle =\sum\limits_{n}{k_{mn}\left\langle E_\text{in}[n] \right\rangle } \propto \sum\limits_{n}{k_{mn}},
\end{equation}
and we use $\mathbf{\Delta E_\textbf{in}}$ to denote the matrix composed of all $\Delta E_\text{in}^s[n]$'s on the right hand side of Eq. \ref{equation8} for simplicity. In the amplitude-only TM method, the output reference light $E_\text{ref}[m]=\sum\limits_{n}k_{mn}$, thus $\left\langle E_\text{out}[m] \right\rangle$ is the same as $E_\text{ref}[m]$ and they serve the same role in providing a benchmark for selecting constructive elements in $K_m$ and blocking the destructive ones. In the amplitude-only TCLF method, since $E_\text{in}[n]$ follows a Bernoulli distribution \cite{zhao2023low}, there is $\left\langle \Delta {{E}_\text{in}}[n]\Delta {{E}_\text{in}}[n] \right\rangle=\frac{1}{4}$. Thus Eq. \ref{equation4} can be written as $\mathbf{\Delta E_\textbf{in}}\cdot {\mathbf{\Delta E_\textbf{in}}}^{T}=\frac{S}{4}\mathbf{I}$, where $\mathbf{I}$ is an identity matrix. Multiplying ${\mathbf{\Delta E_\textbf{in}}}^{T}$ to both sides of Eq. \ref{equation9} and reversing the sides of the equation, we can get
\begin{equation}
    \begin{aligned}
        &\operatorname{Re}\{\left\langle E_\text{out}^{*}[m] \right\rangle\cdot\left[ \begin{matrix}
        {{k}_{m1}} & {{k}_{m2}} &... & {{k}_{mN}}  \\
     \end{matrix} \right]\} = \\ 
       &\quad\frac{4}{S}\operatorname{Re}\{\left\langle E_\text{out}^{*}[m] \right\rangle\cdot\left[ \begin{matrix}
            \Delta E_\text{out}^{1}[m]  & ... & \Delta E_\text{out}^{S}[m]  \\
         \end{matrix} \right]\}\cdot {\mathbf{\Delta E_\textbf{in}}}^{T}. \\ 
    \end{aligned} 
    \label{equation11}      
\end{equation}
In the amplitude-only TCLF method, the two terms on the right side of Eq. \ref{equation5} are conjugate to each other, so it can be written as
\begin{equation}
    \begin{aligned}
        &\left\langle \Delta {{E}_{\text{in}}}[n]\Delta {{I}_\text{out}}[m] \right\rangle \\& \qquad =\left\langle \Delta {{E}_\text{in}}[n]\Delta {E}_\text{in}[n] \right\rangle 2\operatorname{Re}\{\left\langle {{E}^{*}_\text{out}}[m] \right\rangle {{k}_{mn}}\} \\
        &\qquad=2\operatorname{Re}\{\left\langle E^*_\text{out}[m] \right\rangle\left\langle \Delta {{E}_\text{in}}[n]\Delta {{E}_\text{out}}[m] \right\rangle \}.
        \label{equation12}
    \end{aligned}
\end{equation}
This equation reveals another way to characterize the real part of a light field with intensity information other than Eq. \ref{equation7}. The right side of Eq. \ref{equation11} can be replaced by the right side of Eq. \ref{equation12} since they both contain the second-order correlation of the fluctuations of the input and output light fields, namely $\left\langle \Delta {{E}_\text{in}}[n]\Delta {{E}_\text{out}}[m] \right\rangle$. Thus we can substitute $\left\langle \Delta {{E}_{\text{in}}}[n]\Delta {{I}_\text{out}}[m] \right\rangle$ to the right hand side of Eq. \ref{equation11}, and the real part of the rotated $\Delta E_\text{out}^s[m]$ in Eq. \ref{equation11} will be alternated with the output light intensities $\Delta I_\text{out}^s[m]$, as shown below
\begin{equation}
    \begin{aligned}
        &\operatorname{Re}\{\left\langle E_\text{out}^{*}[m] \right\rangle\cdot\left[ \begin{matrix}
        {{k}_{m1}} & {{k}_{m2}} &... & {{k}_{mN}}  \\
     \end{matrix} \right]\} = \\ 
       &\qquad \qquad \qquad \frac{2}{S}\left[ \begin{matrix}
            \Delta I_\text{out}^{1}[m]  & ... & \Delta I_\text{out}^{S}[m]  \\
         \end{matrix} \right]\cdot {\mathbf{\Delta E_\textbf{in}}}^{T}. \\ 
    \end{aligned} 
    \label{equation13}      
\end{equation}
Since $\left\langle \Delta {{E}_{\text{in}}}[n]\right\rangle=0$, we can change the TCLF expression slightly
\begin{equation}
    \begin{aligned}
         \left\langle \Delta {{E}_{\text{in}}}[n]\Delta {{I}_\text{out}}[m] \right\rangle & = \left\langle \Delta {{E}_{\text{in}}}[n]\cdot\{{{I}_\text{out}}[m] - \left\langle {{I}_\text{out}}[m]\right\rangle \}\right\rangle \\
         & = \left\langle \Delta {{E}_{\text{in}}}[n] {{I}_\text{out}}[m]\right\rangle,
    \label{equation13.1} 
    \end{aligned}
\end{equation}
Thus Eq. \ref{equation13} can also be written as
\begin{equation}
    \begin{aligned}
        &\operatorname{Re}\{\left\langle E_\text{out}^{*}[m] \right\rangle \cdot K_m\} = \frac{2}{S}\left[ \begin{matrix}
            I_\text{out}^{1}[m]  & ... & I_\text{out}^{S}[m]  \\
         \end{matrix} \right]\cdot {\mathbf{\Delta E_\textbf{in}}}^{T}. \\ 
    \end{aligned} 
    \label{equation13.2}      
\end{equation}
After recording $I_\text{out}^s[m]$ with the CMOS, the rotated real part of elements in $K_m$ can be calculated by Eq. \ref{equation13.2}. The constructive and destructive elements are separated by 0, with constructive elements being above zero. By contrast, in the amplitude-only TM method, since the conversion from the real part of light fields to intensities is an approximation, the threshold that distinguishes the constructive and destructive elements is manually selected to ensure that the quantities of the two parts are consistent (the threshold is very close to zero in simulations). It is also worth noting that Eq. \ref{equation13.2} shares the same structure as Eq. \ref{equation7.1}: ${\mathbf{\Delta E_\textbf{in}}}$ corresponds to the Hadamard matrix, while $I_\text{out}^{s}[m]$ corresponds to $I_\text{RH}^{s}$. If we inspect these two methods from the perspective of the experimental procedures and the computational processes to generate the TCLF and the binary TM pattern, while omitting their different underlying theories, the only difference between them is the use of the Hadamard matrix versus ${\mathbf{\Delta E_\textbf{in}}}$. Therefore, the orthogonal matrix applied in the amplitude-only TM method doesn't have to be a Hadamard matrix; an "orthogonal" random matrix, namely ${\mathbf{\Delta E_\textbf{in}}}$, would suffice. 

\begin{figure}[htbp]
\centering
\includegraphics[width=\linewidth]{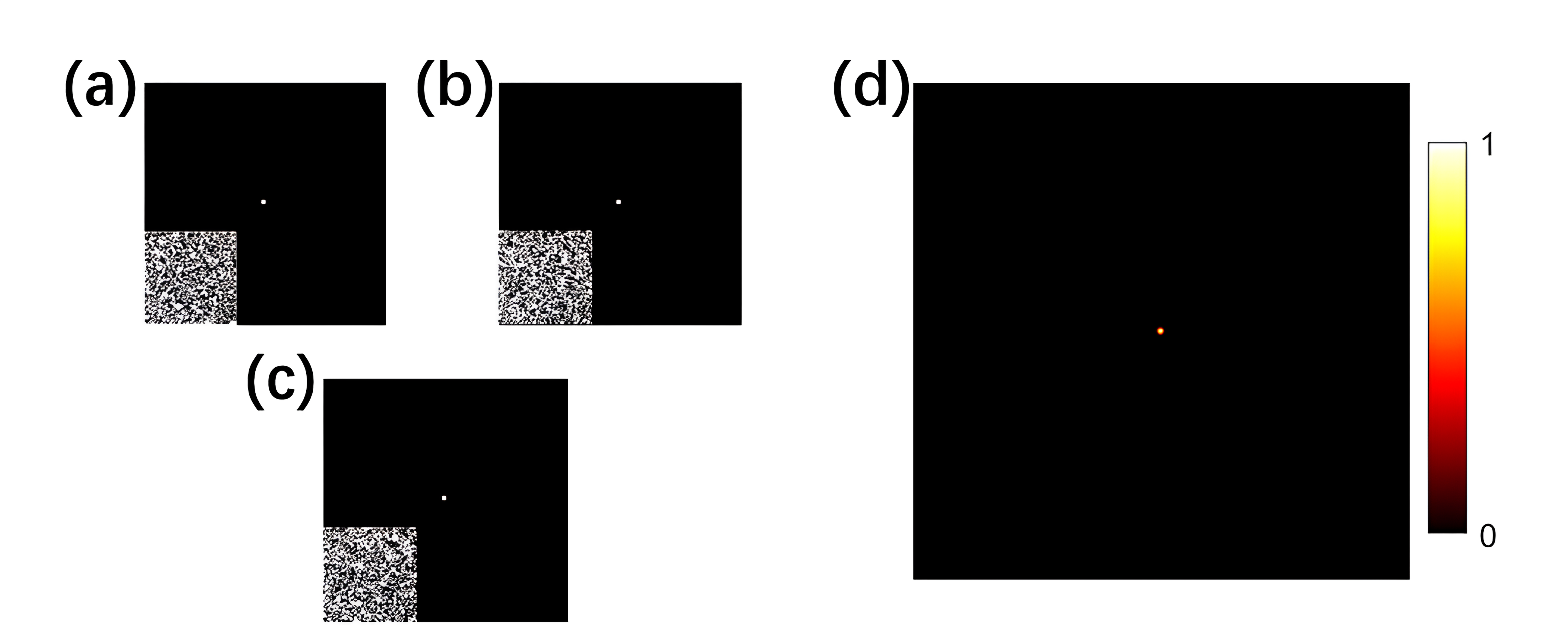}
\caption{(a)Numerical simulation of the amplitude-only TCLF pattern (in the bottom left corner) and the focus formed on the CMOS plane when reloading it onto the DMD in the amplitude-only TCLF method. (b)Numerical simulation of the binary TM pattern and the focus in the amplitude-only TM method. (c)Numerical simulation of the trial where the Hadamard matrix is replaced with $\mathbf{\Delta E_\textbf{in}}$ while keeping other parts of the amplitude-only TM method unchanged. (d)Experimental result of the trial.}
\label{picture2}
\end{figure}
A simple numerical simulation is performed to verify the above conclusions before we explain the reason for the different loads on PC memory when applying these two methods and give some comments on the relationship between the Hadamard matrix and ${\mathbf{\Delta E_\textbf{in}}}$. We first verify that these two amplitude-only methods yield similar results and the binary TM pattern is basically the same as the TCLF pattern. The simulation configuration is shown in Fig. \ref{picture1}. A laser beam with a wavelength of $\lambda=632.8nm$ is incident on a DMD with an area of $3mm\times3mm$ (we only use $64\times64$ pixels). The scattering object is simulated by generating a 2D array with random phases. The distance between the DMD and the scattering object is $z_1=300mm$ and the distance between the scattering object and the CMOS camera is $z_2=400mm$. We use the Fresnel diffraction formula to simulate light propagation. The amplitude-only TM simulation is performed by loading Hadamard bases onto the DMD along with the reference light and measuring the corresponding light intensities on the CMOS plane. The amplitude-only TCLF simulation is performed with $S=2\times 10^4$ sets of random input patterns that follow the Bernoulli distribution. After computing the binary TM and TCLF pattern, they are reloaded on the DMD to form a focus respectively, as shown in Fig. \ref{picture2}(a) and \ref{picture2}(b). We use the enhancement of the focus $\eta ={{I}_{\text{focus}}}/{{I}_\text{background}}$ to evaluate focusing quality, where ${{I}_{\text{focus}}}$ represents the intensity of the focus and ${{I}_\text{background}}$ denotes the average intensity in other areas.
\begin{figure}[htbp]
\centering
\includegraphics[width=\linewidth]{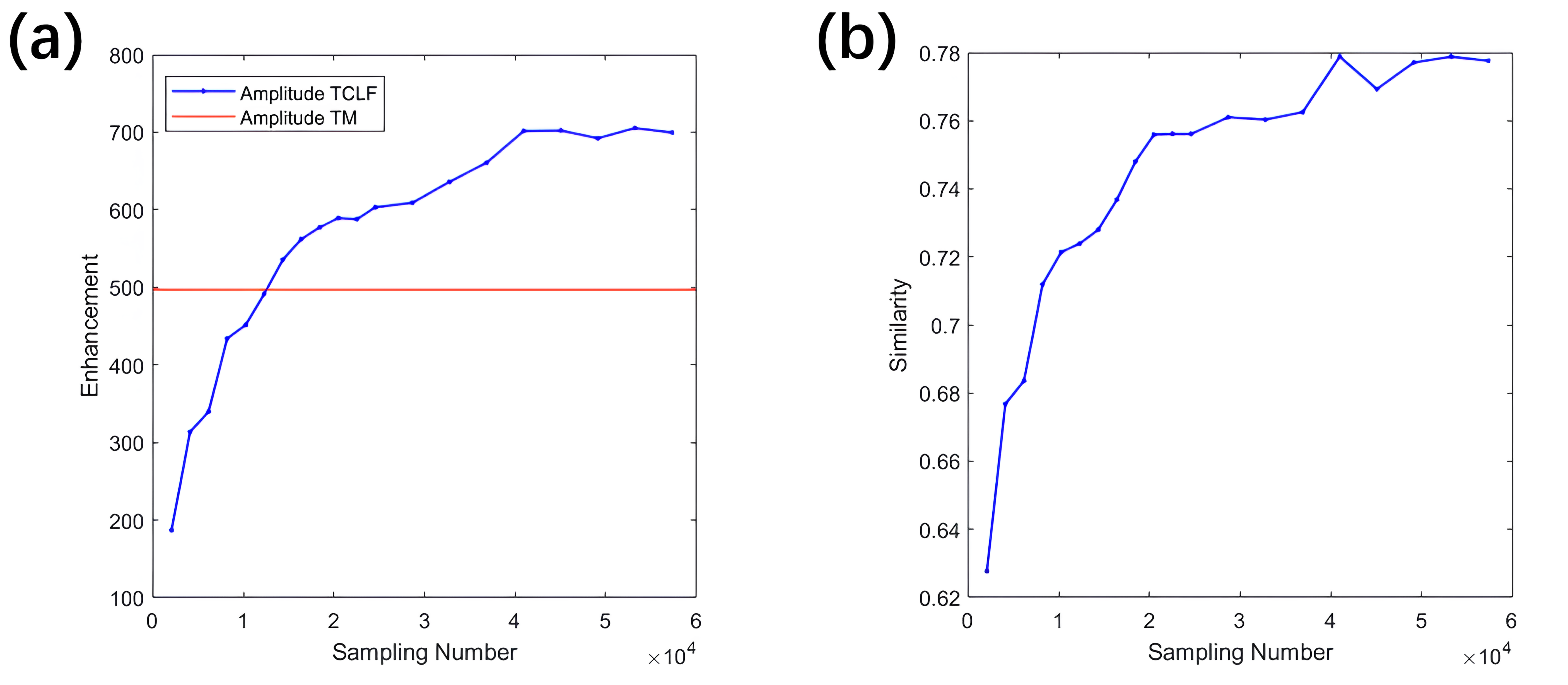}
\caption{(a)The blue curve is the relationship between the focus enhancement and the sampling number in the amplitude-only TCLF method, and the red curve is the focus enhancement of the amplitude-only TM method. (b)The similarity between the binary TM and the TCLF pattern versus the sampling number.}
\label{picture3}
\end{figure}
In Fig. \ref{picture3}(a), the maximum focus enhancement for the amplitude-only TCLF is $701$, while the focus enhancement is $497$ for the amplitude-only TM method. We define the similarity between the binary TM and the TCLF pattern by the proportion of identical elements to the total number of elements. The maximum similarity is about $78\%$, as shown in Fig. \ref{picture3}(b). It can reach $85.7\%$ if we use $32 \times 32$ pixels on the DMD. Note that the sampling number of the amplitude-only TCLF method $S$ is variable. The focus enhancement and the similarity are proportional to the sampling number $S$, with their peaks at about $S=4\times 10^4$. 

We mentioned the only difference between these two methods is the use of the Hadamard matrix and ${\mathbf{\Delta E_\textbf{in}}}$. We test this idea by replacing the Hadamard matrix with $\mathbf{\Delta E_\textbf{in}}$ in the amplitude-only TM simulation while maintaining other parts unchanged. The simulation shows that a clear focus can form on the CMOS camera plane, as shown in Fig. \ref{picture2}(c). The same result can also be drawn from the experiment that performs the test, as shown in Fig. \ref{picture2}(d). The experimental setup is the same as the one in the paper \cite{zhao2023low} that proposed the TCLF method, with $ 500\times500$ pixels used on the DMD and $S=4\times 10^5$. As we demonstrate above the amplitude-only TCLF can be explained in the paradigm of the amplitude-only TM method, it may work the other way around as well: the Hadamard bases are samples of the random vector $\Delta E_\text{in}^{s}$ that consists of uniformly distributed 1s and -1s, except for the first column of the Hadamard matrix; thus the Hadamard matrix can be treated as a special case of the "orthogonal" random matrix ${\mathbf{\Delta E_\textbf{in}}}$ (${\mathbf{\Delta E_\textbf{in}}}$ won't be strictly orthogonal unless $S$ is infinitely large) and the amplitude-only TM is a TCLF method in essence. Note that the theory of the amplitude-only TM method contains approximations in Eq. \ref{equation7} and \ref{equation7.1}, while the derivation of the TCLF method from Eq. \ref{equation8} to Eq. \ref{equation13.2} is rigorous and has clear physical and statistical significance. Therefore, it may be more appropriate to consider the Hadamard matrix as a sample of ${\mathbf{\Delta E_\textbf{in}}}$ and attribute the success of the "amplitude-only TM method" to the TCLF theory that utilizes statistical properties of random fluctuations of the light source, rather than the traditional TM theory that uses orthogonal bases with fixed values. 
%As we demonstrate above the amplitude-only TCLF can be explained in the paradigm of the amplitude-only TM method, it may work the other way around as well: the Hadamard bases are samples of the random vector $\Delta E_\text{in}^{s}$ except for the first column of the Hadamard matrix, thus the Hadamard matrix can be treated as a special case of the "orthogonal" random matrix ${\mathbf{\Delta E_\textbf{in}}}$ and the amplitude-only TM is a TCLF method in essence. Note that the theory of the amplitude-only TM method contains approximations in Eq. \ref{equation7} and \ref{equation7.1}, while the derivation in the TCLF theory is rigorous and has clear physical significance. Therefore, it may be more appropriate to consider the Hadamard matrix as a sample of ${\mathbf{\Delta E_\textbf{in}}}$ and attribute the experimental success of the "amplitude-only TM method" to the TCLF theory, rather than the TM theory.

Now we can explain the reason for the significantly different loads on PC memory when applying them. If we use $1,024\times 1,024$ pixels on the DMD for the amplitude-only TM method, generating the Hadamard matrix ($1,024^2 \times 1,024^2$) would occupy $2TB$ of memory. This issue becomes more pronounced as the matrix size increases. In another research \cite{yu2017ultrahigh}, the Hadamard matrix is divided into several sub-matrices to mitigate the heavy load. By contrast, the amplitude-only TCLF method, which employs an "orthogonal" random matrix $\mathbf{\Delta E_\textbf{in}}$, avoids the need for generating such an orthogonal matrix. Instead, it generates a column of random light fields and deletes them after recording the corresponding $I_\text{out}^s$ with the CMOS at each sampling. This ensures that only $1,024\times 1,024$ data are stored in memory at any time, which consumes only $2MB$.

In conclusion, our findings reveal a fundamental connection between the amplitude-only TM and amplitude-only TCLF methods.  However, it has to be acknowledged that this theory only works for the amplitude-only methods and does not extend to the phase-only methods. It appears that the phase-only TM method shares no deep connection with the phase-only TCLF method \cite{popoff2009measuring}. Further exploration is needed to fully elucidate their relationship.

\begin{backmatter}
%\bmsection{Funding} Content in the funding section will be generated entirely from details submitted to Prism. Authors may add placeholder text in the manuscript to assess length, but any text added to this section in the manuscript will be replaced during production and will display official funder names along with any grant numbers provided. If additional details about a funder are required, they may be added to the Acknowledgments, even if this duplicates information in the funding section. See the example below in Acknowledgements. For preprint submissions, please include funder names and grant numbers in the manuscript.

%\bmsection{Acknowledgments} The section title should not follow the numbering scheme of the body of the paper. Additional information crediting individuals who contributed to the work being reported, clarifying who received funding from a particular source, or other information that does not fit the criteria for the funding block may also be included; for example, ``K. Flockhart thanks the National Science Foundation for help identifying collaborators for this work.''

\bmsection{Disclosures} The authors declare no conflicts of interest.

\iffalse
\smallskip

\noindent Here are examples of disclosures:

\bmsection{Disclosures} ABC: 123 Corporation (I,E,P), DEF: 456 Corporation (R,S). GHI: 789 Corporation (C).

\bmsection{Disclosures} The authors declare no conflicts of interest.
\fi

\bmsection{Data Availability Statement} The data that support the findings of this study are available from the corresponding author upon reasonable request.

\bigskip

%\noindent Data availability statements are not required for preprint submissions.

%\bmsection{Supplemental document}
%See Supplement 1 for supporting content. 

\end{backmatter}

%\section{References}

%Note that \emph{Optics Letters} and \emph{Optica} short articles use an abbreviated reference style. Citations to journal articles should omit the article title and final page number; this abbreviated reference style is produced automatically when the \emph{Optics Letters} journal option is selected in the template, if you are using a .bib file for your references.

%\bigskip
%\noindent Add citations manually or use BibTeX. See \cite{Zhang:14,OPTICA,FORSTER2007,testthesis,manga_rao_single_2007}. List up to three author names in references, and if there are more than three authors use \emph{et al.} after that.

% Bibliography
\bibliography{sample}

% Full bibliography added automatically for Optics Letters submissions; the following line will simply be ignored if submitting to other journals.
% Note that this extra page will not count against page length
\bibliographyfullrefs{sample}

%Manual citation list
%\begin{thebibliography}{1}
%\bibitem{Zhang:14}
%Y.~Zhang, S.~Qiao, L.~Sun, Q.~W. Shi, W.~Huang, %L.~Li, and Z.~Yang,
 % \enquote{Photoinduced active terahertz metamaterials with nanostructured
  %vanadium dioxide film deposited by sol-gel method,} Opt. Express \textbf{22},
  %11070--11078 (2014).
%\end{thebibliography}

% Please include bios and photos of all authors for aop articles
\ifthenelse{\equal{\journalref}{aop}}{%
\section*{Author Biographies}
\begingroup
\setlength\intextsep{0pt}
\begin{minipage}[t][6.3cm][t]{1.0\textwidth} % Adjust height [6.3cm] as required for separation of bio photos.
  \begin{wrapfigure}{L}{0.25\textwidth}
    \includegraphics[width=0.25\textwidth]{john_smith.eps}
  \end{wrapfigure}
  \noindent
  {\bfseries John Smith} received his BSc (Mathematics) in 2000 from The University of Maryland. His research interests include lasers and optics.
\end{minipage}
\begin{minipage}{1.0\textwidth}
  \begin{wrapfigure}{L}{0.25\textwidth}
    \includegraphics[width=0.25\textwidth]{alice_smith.eps}
  \end{wrapfigure}
  \noindent
  {\bfseries Alice Smith} also received her BSc (Mathematics) in 2000 from The University of Maryland. Her research interests also include lasers and optics.
\end{minipage}
\endgroup
}{}

\end{document}